\documentclass[conference]{IEEEtran}
\pdfoutput=1

\usepackage{cite}

\ifCLASSINFOpdf
  \usepackage[pdftex]{graphicx}
  \graphicspath{{../pdf/}{../jpeg/}}
  \DeclareGraphicsExtensions{.pdf,.jpeg,.png}
\else
\fi

\ifCLASSOPTIONcompsoc
    \usepackage[caption=false,font=normalsize,labelfont=sf,textfont=sf]{subfig}
\else
    \usepackage[caption=false,font=footnotesize]{subfig}
\fi

\usepackage{scalerel}
\usepackage{tikz}
\usetikzlibrary{svg.path}
\definecolor{orcidlogocol}{HTML}{A6CE39}
\tikzset{
  orcidlogo/.pic={
    \fill[orcidlogocol] svg{M256,128c0,70.7-57.3,128-128,128C57.3,256,0,198.7,0,128C0,57.3,57.3,0,128,0C198.7,0,256,57.3,256,128z};
    \fill[white] svg{M86.3,186.2H70.9V79.1h15.4v48.4V186.2z}
                 svg{M108.9,79.1h41.6c39.6,0,57,28.3,57,53.6c0,27.5-21.5,53.6-56.8,53.6h-41.8V79.1z M124.3,172.4h24.5c34.9,0,42.9-26.5,42.9-39.7c0-21.5-13.7-39.7-43.7-39.7h-23.7V172.4z}
                 svg{M88.7,56.8c0,5.5-4.5,10.1-10.1,10.1c-5.6,0-10.1-4.6-10.1-10.1c0-5.6,4.5-10.1,10.1-10.1C84.2,46.7,88.7,51.3,88.7,56.8z};
  }
}
\newcommand\orcidicon[1]{\href{https://orcid.org/#1}{\mbox{\scalerel*{
\begin{tikzpicture}[yscale=-1,transform shape]
\pic{orcidlogo};
\end{tikzpicture}
}{|}}}}

\usepackage{url}
\usepackage{color} %

\usepackage[hidelinks]{hyperref}

\hypersetup{
pdftitle={A Concept for a Qualifiable (Meta)-Modeling Framework Deployable in Systems and Tools of Safety-critical and Cyber-physical Environments},
pdfsubject={},
pdfauthor={Vanessa Tietz; Julian Schoepf; Andreas Waldvogel; Bjoern Annighoefer},
pdfkeywords={Ada SPARK, domain-specific modeling, (meta)-modeling framework, model-based systems, engineering, model transformation language, qualifiable runtime environment, safety-critical, (tool) qualification, visualization}
}

 \makeatletter
 \newcommand{\linebreakand}{%
   \end{@IEEEauthorhalign}
   \hfill\mbox{}\par
   \mbox{}\hfill\begin{@IEEEauthorhalign}
 }
 \makeatother

\newcommand{\includevisio}[2][]{\includegraphics[clip, trim=0.2cm 0.2cm 0.3cm 0.3cm, #1]{#2}} 

\newcommand\copyrighttext{%
  \footnotesize \textcopyright 2021 IEEE. Personal use of this material is permitted.
  Permission from IEEE must be obtained for all other uses, in any current or future
  media, including reprinting/republishing this material for advertising or promotional
  purposes, creating new collective works, for resale or redistribution to servers or
  lists, or reuse of any copyrighted component of this work in other works.
  }
\newcommand\copyrightnotice{%
\begin{tikzpicture}[remember picture,overlay]
\node[anchor=south,yshift=10pt] at (current page.south) {\fbox{\parbox{\dimexpr\textwidth-\fboxsep-\fboxrule\relax}{\copyrighttext}}};
\end{tikzpicture}%
}

\begin{document}
\title{A Concept for a Qualifiable (Meta)-Modeling Framework Deployable in Systems and Tools of Safety-critical and Cyber-physical Environments}

\author{\IEEEauthorblockN{Vanessa Tietz \orcidicon{0000-0002-5942-5893}}
\IEEEauthorblockA{Institute of Aircraft Systems\\
University of Stuttgart\\
Stuttgart, Germany\\
0000-0002-5942-5893}
\and
\IEEEauthorblockN{Julian Schoepf \orcidicon{0000-0003-1110-2830}}
\IEEEauthorblockA{Institute of Aircraft Systems\\
University of Stuttgart\\
Stuttgart, Germany\\
0000-0003-1110-2830}
\and
\IEEEauthorblockN{Andreas Waldvogel \orcidicon{0000-0003-4819-9031}}
\IEEEauthorblockA{Institute of Aircraft Systems\\
University of Stuttgart\\
Stuttgart, Germany\\
0000-0003-4819-9031}
\linebreakand
\IEEEauthorblockN{Bjoern Annighoefer \orcidicon{0000-0002-1268-0862}}
\IEEEauthorblockA{Institute of Aircraft Systems\\
University of Stuttgart\\
Stuttgart, Germany\\
0000-0002-1268-0862}}

\maketitle
\copyrightnotice
\vspace*{-0.86\baselineskip} %

\begin{abstract} 
The development of cyber-physical systems can significantly benefit from domain-specific modeling and requires adequate (meta)-modeling frameworks. If such systems are designed for the safety-critical area, the systems must undergo qualification processes defined and monitored by a certification authority. 
To use the resulting artifacts of modeling tools without further qualification activities, the modeling tool must be additionally qualified. Tool qualification has to be conducted by the tool user and can be assisted by the tool developer by providing qualification artifacts. 
However, state-of-the-art domain-specific modeling frameworks barely support the user in the qualification process, which results in an extensive manual effort. 
To reduce this effort and to avoid modeling constructs that can hardly be implemented in a qualifiable way, we propose the development of an open source (meta)-modeling framework that inherently considers qualification issues. 
Based on the functionality of existing frameworks, we have identified components that necessarily need to be rethought or changed. This leads to the consideration of the following six cornerstones for our framework: (1) an essential meta-language, (2) a minimal runtime, (3) deterministic transformations, (4) a qualification artifact generation, (5) a sophisticated visualization, and (6) a decoupled interaction of framework components. All these cornerstones consider the aspect of a safety-critical (meta)-modeling framework in their own manner. This combination leads to a holistic framework usable in the safety-critical system development domain. 

\end{abstract}

\IEEEpeerreviewmaketitle

\section{Introduction}
Domain-specific modeling (DSM) has attracted popularity for system developers for several years. Multiple meta-languages and modeling frameworks exist to address numerous use cases.
DSM easily enables working with higher level system abstractions and reusability as well as efficient automated processes with model transformations. In addition, handling of complex and abstract systems can be supported by graphical visualization and editing.
Since common DSM frameworks do not focus on the application in safety-critical domains, there is no need to support tool qualification, even if qualifiability has been identified as a key element for modeling tools \cite{Paige2012}. In practice, DSM often only extends auxiliary tools. If the tool user nevertheless wants to use common frameworks in the safety-critical domain, this leads to high qualification effort. This is due to fragmentary or nontransparent information on the background and structure of the used tools. Moreover, one problem is the strong dependence between framework components and the custom models. 
For the application of DSM in safety-critical systems with high demands towards qualifiability, we propose an essential (meta)-modeling and transformation framework. The framework focuses on the efficient qualification and automation of processes for safety-critical domains like automotive or aerospace. This is achieved by introducing a meta-language considering qualification aspects, implementing the meta-language with the formal provable but strongly restrictive programming language Ada, performing transformations deterministically and traceably, visualizing complex models, and supporting the generation of qualification artifacts.
Another key property is a strong decoupling of model, runtime, and user interface. Overall, the framework shall be a new chance where state-of-the-art tools do not suit.

The remainder of this paper is organized as follows. The next section addresses foundations. The third section reviews the most recent work over the last decades regarding the modeling of safety-critical systems, transformations, and visualization. Our vision is presented in section four, while the first progress of our new framework is described in section five. Section six gives a short conclusion with further research tasks. 

\section{Nomenclature, Multi-level Modeling, Visualization and Qualification}
This section provides a consistent view of our framework by introducing the nomenclature, subsequently describing foundations of used methods. For our framework, we do not just rely on one existing meta-modeling standard. Where needed, the advantages of several standards and concepts are combined.  Therefore, multilevel-modeling and deep instantiation are explained followed by existing basics of model-to-model transformations, generic visualization, and tool qualification.

\subsection{Nomenclature} 
 In (meta)-modeling, different layers have to be distinguished. In the following, we use the designation from the meta-object-facility (MOF) standard \cite{MOF}: The real world is the M0-level, the domain-specific model the M1-level, the meta-model the M2-level, and the meta-language the M3-level.
 
\subsection{Multi-level Modeling}
The modeling of systems and software is often performed with the Unified Modeling Language (UML)\cite{UML}. This leads to the typical structure of two interacting levels - a classification level (UML-Language) and an instantiation level (UML-Model). Modeling with more than two levels is called multi-level modeling. The principle of instantiating the lower level remains. Multi-level modeling can help to overcome problems regarding the explicit specification of relationships with UML that are otherwise only assumed by convention. One multi-level artifact addresses logical and physical classification. That concept is used to give every element on every modeling level the meaning of a physical object \cite{DeepInstantiation}. The concept is illustrated with a dashed line and the annotation 'physical' in diagrams.

A second artifact is that in the UML standard common elements defined on a modeling level can only be instantiated at the level above - the 'two-levels only' modeling philosophy. The problem of instantiating objects more than one level above can be addressed with deep instantiation \cite{DeepInstantiation}, which introduces a potency value that allows to define in how many levels an instantiation shall be possible. Every time the element is instantiated, the potency value is decreased by one. When reaching a potency value of zero, no more instantiations are possible. Deep instantiation can be visualized with the '@' symbol or super-scripted numbers. This concept of deep instantiation is based on the assumption that each element in multi-level modeling can act simultaneously as class and object.

\subsection{Automation via Model-to-Model Transformation (M2M)}
Model transformations enable reusable processing of domain-specific models. For the implementation of such transformations, multiple approaches with specific characteristics exist. In \cite{Biehl} different classification schemes for transformations are given. Transformation languages can be distinguished through different language paradigms, e.g. relational (TXL \cite{TXL}), imperative (KERMETA \cite{KERMETA}), graph-based (GReAT \cite{GREAT}), or hybrid languages (ATL \cite{ATL}). Properties like rule application control, scheduling, and organization are important aspects, which affect the termination, processing behavior (e.g. in parallel or recursively), and reusability of the transformations rules. There are more specific characteristics described in \cite{Biehl}, \cite{DiRuscio}, and \cite{Kahani2018SurveyAC} of model transformations all with different advantages and disadvantages for specific problem domains. 

\subsection{Generic Visualization of Models}
Generally, domain-specific models are visualized interactively to facilitate understanding and manipulation. Since users mainly interact with the visualization, it has a strong effect on the user experience. Among the four classes of visualization techniques in \cite{Keim1996}, the icon-based technique class is dominant in modeling. 
To visualize models 2-dimensionally, graphs are typically used. For instance, objects are visualized by vertices and references by edges. Well established is the software design pattern model-view-controller \cite{Leff2001}. 

Visualization elements can be directly defined in the meta-model or they are decoupled and provided by a separate presentation language \cite{Zhang2015}. The latter has the advantage of keeping the meta-model simple, which can be used to reduce qualification efforts. Generic visualization can be used in both cases and adapts the visualization of the DSM environment dynamically to the meta-model, e.g. only valid child elements are made available in a palette \cite{UniISISVanderbilt2017}.

\subsection{Tool Qualification}
For the development of cyber-physical systems, tools can assist development and verification to decrease manual workload and costs. When developing safety-critical systems, it is necessary to avoid negative impact of error-prone tools on the system functionality \cite{DO330}. This can be achieved by verifying the generated artifacts of used software and tools manually or by using qualified tools. There are different standards considering the process of tool qualification. For instance, ISO-26262 \cite{ISO26262} and DO-330 \cite{DO330} define tool qualification in the automotive industry and aviation, respectively. Tools can be divided into different tool qualification levels (TQLs) based on their intended use and safety level. Depending on the qualification level, different requirements and processes have to be fulfilled in order to obtain a qualified tool. Qualifiable in this context means that a system must be successfully validatable against the relevant standards and regulations and finally acceptable to a certification authority. The tool user has full responsibility for the tool qualification. Moreover, the qualification is bound to a specific application \cite{ibrahim2021state}. The tool developer or supplier can support tool qualification by providing artifacts demanded by tool qualification standards, i.e. a tool qualification kit. Qualification kits may include the description of the tool functions, technical features, user instructions, list of error-messages and the usage domain, data flow and control flow, requirements, and test reports as required in ISO-26262 or DO-330.

\section{Related Work and Limitations}
DSM is a method in system and software engineering, which is used in a variety of different applications under usage of different frameworks e.g. Eclipse Modeling Framework (EMF) \cite{EMF} or Generic Modeling Environment (GME) \cite{GME}. Most of these frameworks are based on the Meta Object Facility (MOF)\cite{MOF} or the Unified Modeling Language (UML)\cite{UML} standard. 
Their high complexity and missing artifacts to support qualification impede the use of such tools for the development of safety-critical systems. In addition, they often contain modeling elements that can hardly be programmed deterministically, e.g. proxies, opposite references, or object-oriented implementations.

In recent years, there have been attempts to make modeling frameworks and tools easier to use for the development of safety-critical systems. Mainly, the automatic generation of qualification artifacts is used for this purpose. In \cite{proll2017applying}, qualification artifacts can be generated with model-to-text (M2T) transformations from the combination of different domain-specific languages (DSL) (e.g. system structure, or requirements). All these DSLs are mainly based on the meta-languages Ecore and UML. An EMF-based tool for modeling and code generation of safety-critical systems is proposed in \cite{barth2019modeling}, \cite{barthmodeling}. Again, the focus is on the automatic generation of qualification artifacts. In addition, emphasis is placed on an easily readable as well as qualifiable generated code. An own simple and easy to use safety domain-specific language based on the Ecore meta-language was developed. With their own DSL, different system views (safety requirements, functional view, hardware view, and software view) are modeled and then combined. With the resulting model, it is possible to generate code files and safety documentation. 
Apart from the implementations already mentioned, we are unaware of any other approaches targeting the issue of qualifiable safety-critical system modeling. 
It is important to note that our framework alone will not be a safety-critical product. Therefore, approaches doing qualification processes are out of scope. The framework shall facilitate qualification processes for the user, but without impact on the conducted processes.

Considering model transformation in the safety-critical domain, the following standards and works have to be mentioned.  A well-known standard for model transformation is Query/View/Transformation (QVT) \cite{QVT}, which serves as a basis for numerous model transformation approaches. 
However, many transformation languages do not consider qualification purposes like deterministic transformation behavior. Moreover, traceability and debugging features are often considered within the concepts of the languages, but must be addressed separately. 
As with the modeling of safety-critical systems, it is possible to use state-of-the-art transformation languages and check their artifacts either manually or automatically. For instance, in safety-critical avionics GReAT is used for automatic generation of configuration artifacts \cite{BelKra}, as well as for the generation of requirement documents \cite{Belschner} and the dedicated test artifacts \cite{Mueller} \cite{AAA}. However, the output has to be validated manually because the correctness of the transformation is not ensured. For verifying the correctness of model transformations, in the last decade, many approaches were examined and developed. A summary can be found in \cite{VerifyingMT}. The focus of these approaches relies on different aspects like testing, theorem proving, or model checking. However, we are unaware of transformation languages that are implemented in a deterministic manner with a qualifiable programming language.

The use of web browsers for visualization is becoming increasingly popular due to their native support for HTML5, JavaScript, and SVG. These web applications combine the benefits of desktop applications like speed, single-page view, and context menus with the advantages of cloud-based solutions like simple deployment, ensured updates, and the business model software-as-a-service. This trend towards browser-based solutions has reached DSM as well \cite{Gray2016}. 
Recent work targets web-based collaborative meta modeling, e.g. emf.cloud \cite{Foundation2020}, WebGME \cite{Maroti2014}, and AToMPM \cite{Syriani2013}.
Diagram rendering and interaction, in this case, is based on JavaScript libraries like mxGraph \cite{JGraph2020}, JointJs \cite{IO2021}, D3.js \cite{Bostock2011}, and Eclipse Sprotty \cite{EclipseFoundation2019}.
In no solution, we see a clear separation of diagrams and data, such that visualization and editing are subject to qualification, including diagram libraries. Moreover, many frameworks and also diagram libraries do not support complex system data, e.g. the interactive modeling of nested signal flows. 
Scalability has been identified as a key element for graphical editing \cite{Kolovos2013}. Handling complex models is typically addressed by high-performance rendering and layouting libraries in combination with information restriction, e.g. by user-defined views. For instance, WebGME provides crosscuts, which allow visualizing chosen associations across multiple model hierarchy levels \cite{Maroti2014}. Further functionalities that help the user to understand or modify the model as efficiently and conveniently as possible are dynamic level of detail \cite{Nugroho2009}, connectivity abstraction \cite{Zhang2015}, dynamic highlighting \cite{itemis2021}, dynamic filtering \cite{Viyovic2014}, and auto-layout \cite{Eiglsperger2004}.

In summary, existing frameworks and works show the benefit of meta-modeling and derived tools for the cyber-physical domain. However, the industrial utilization of such tools for developing safety-critical systems is rare. In our opinion, this is caused by 1) non-deterministic and traceable behavior, 2) non-limited scope of functionality, 3) unfeasible or difficult qualification of used programming language, 4) complex dependencies of meta-modeling, runtime, visualization and persistence, 5) missing tool qualification artifacts or their deployment for the user. Moreover, 6) user experience is degraded by overloaded user interfaces, inappropriate visualizations, and slower editing than textual file modifications. Last, 7) most existing frameworks do not offer sophisticated debugging functionality, which is assumed important for tool and automation development. 

\section{Six Cornerstones of Qualifiable Domain-specific Modeling} 
Our proposed domain-specific modeling framework that supports qualification and solves described issues consists of six cornerstones: \textit{(A) An Essential Meta-Language:} Existing meta-languages are complex and often contain modeling constructs that are difficult to qualify. \textit{(B) A Minimalistic, Qualifiable Runtime:} Meta-languages and transformations are often implemented in hardly qualifiable programming languages, implementing ours with Ada might be a solution. \textit{(C) Deterministic, Verifiable, (Graphical) Transformations:} The aim is to develop a deterministic and traceable transformation language. \textit{(D) Qualification Artifact Generation:} To support the documentation and qualification of generated models and tools, qualification artifacts as requirements and tests are required. \textit{(E) Decoupled Visualization:} For the visualization and the interaction with complex system models, a sophisticated visualization environment is desired that shall not impede qualification. 
\textit{(F) Decoupled Interaction of Framework Components:} To qualify the cornerstones separately, maximum achievable modularity and independence of the cornerstone implementations are desired. 
In the following, each of the cornerstones is described in detail.

\subsection{An Essential Meta-Language}
State-of-the-art meta-languages provide a huge range of modeling possibilities in order to be usable in a lot of different applications, e.g. data models, but also software code generation. The goal in developing our meta-language is to provide essential and simplified functionality to meet essential needs in modeling purely static elements, attributes, and relations in (safety-critical) systems. The fewer elements there are in the meta-language, the fewer functions and code need to be validated and the fewer (negative) side effects can be produced by the tool user. Of course, an essential functionality cannot cover every possible application of common languages. In addition to the essential functionality, modeling constructs that are difficult to implement in a deterministic manner like opposite references or proxies shall be avoided. For instance, proxies can be implemented with the use of pointers which is not advisable from a safe programming perspective \cite{MISRAC}. The same applies to opposite references.  The intention is explicitly not to eliminate all elements that are difficult to qualify, but to create alternatives where necessary.
The goal is not only to focus on essential needs but also to deterministically and reliably implement new concepts like (1) a dedicated constraint environment, or (2) supporting the generation of qualification artifacts from the model. Lastly, it has to be ensured that the created meta-language and its functionalities can be implemented in a way that does not impede qualification of its runtime.

\subsection{A Minimalistic, Qualifiable Runtime}
For the interaction with the developed meta-language, it must be implemented to source code. The target is a minimal code footprint in order to only provide necessary operations to be able to create a meta-model and a domain-specific model from the meta-language and to implement the supplementary functionality described above. Automatic code generation from created models, as is common with other modeling frameworks, will not be provided. Modifications of the M2 and M1 models should be conducted using generic commands and requests. To be able to understand models during runtime, reflective requests are needed. The communication between the runtime and the other cornerstones should only be possible via a generic interface. Accordingly, a suitable interprocess communication is required. In order to ease qualification, the runtime has to be written in a programming language which supports qualification. Therefore, the programming language has to ensure a lot of different objectives listed e.g. in standards comparable to MISRA-C or the DO-178C standard. Requirements are most prominently type-safety, avoidance of dynamic memory allocation (usage of pointers), appropriate error handling, good code readability, deterministic behavior, testability, easy test coverage, absence of unintended functionality, and avoidance of dead code.   

\subsection{Deterministic, Verifiable, (Graphical) Transformations} %
Transformations play an important role in introducing automated workflows for DSM.
Qualifying generated transformation artifacts is a task that needs additional high effort. It is the responsibility of the tool user to prove the correctness of the artifacts, which is difficult if reliable post conditions of the transformation language and engine are not obtained. Therefore, the use of such transformations in safety-critical system development is limited\cite{DO330}.

In our opinion, for an applicable transformation framework, the user has to be supported on different levels: 1) a qualified implementation of the transformation language, 2) deterministic and consistent transformation behavior, 3) integrated  traceability of the results, and 4) debugging features that are decoupled from the qualifiable implementation. These features shall support the user in defining adequate  transformation patterns. The debugging functionality should be additionally switchable on and off. Additionally, 5) a graphical representation for defined transformation rules is desired, because experience showed that graphical representation is more intuitive and easier accessible for users than bare programming code, however the graphical representation is not stated to be mandatory. 

Deterministic behavior means the completion of a transformation in a finite amount of time and the same output for the same input models, e.g. no ambiguity of transformation orders. Moreover, the generated output has to comply with its meta-model structure. Determinism in our opinion is a prerequisite for qualifiable software and therefore is necessary. For example, it is not possible to provide test coverage, if you cannot reach all sections, which is the case for non-deterministic transformations.

\subsection{Qualification Artifact Generation} 
The usability of tools and frameworks in safety-critical system development is strongly connected to the qualification effort of the user. With a specific TQL-level, which is specified in the appropriate standards, it is inherently necessary for the tool user to deliver evidence for the correct operation of its used tools.  The framework should support as much as possible artifacts demanded by tool qualification guidelines, however automating the complete tool qualification process seems infeasible. Moreover, tool qualification has to consider the target application and target environment. Nonetheless, we want to support the user by offering a qualified runtime and an automated generation of qualification artifacts, (e.g. documentation, requirements, and test cases) for defined meta-models and domain-specific models. Additionally, traceability artifacts for model transformation changes shall be generated and if possible also qualification artifacts for them. 
However the task of generating such qualification artifacts is challenging and requires deeper investigations in the future. Overall, the effort of the user for qualifying its application shall be minimized by implementing the above mentioned features.

\subsection{Decoupled Visualization}
Our framework shall fulfill the contradicting requirements of a sophisticated visualization and qualifiability. State-of-the-art visualizations rely on complex libraries and are implemented in  programming languages hard to qualify. This dilemma is overcome by strictly decoupling the visualization from the other components via a qualifiable generic interface. Qualification of the visualization component is therefore not required if it can be shown that the visualization cannot introduce unintended side-effects into the models.
To make the framework instantly available on virtually any device, the framework shall be accessible as a browser-based solution. 
The visualization shall provide an editor for the meta-models, the domain-specific models, and the transformation models, if applicable. Additionally, it shall provide multiple editors for the same model to fulfill the needs of different roles. An editor meta-model facilitates the development of this plurality of editors in a model-driven way.   
The editor shall support the user by instantaneously highlighting inconsistent modeling states that are reported by the runtime. Furthermore, level-of-detail and visual filtering shall be available. 
It is important to note that visual editing is an optional feature for more comprehensive modeling, but all previous components need to be fully operational without visual editing.

\subsection{Decoupled Interaction of Framework Components}
The described cornerstones are decoupled. First, this allows the mixed usage of qualified and not qualified components. Secondly, it facilitates the exchange and independent usage of the components. The decoupling shall be implemented by a standardized interface that provides the functionality of the Object Constraint Language (OCL) \cite{OCL} and in addition transactional model modifications under consideration of the ACID  (Atomicity, Consistency, Isolation, Durability) principle, change events, and error handling. Moreover, all operations conducted with this interface shall behave deterministically. 

\section{Interaction of Cornerstones and First Implementations}
\begin{figure}[!t]
    \centering
    \includevisio[width=0.37\textwidth]{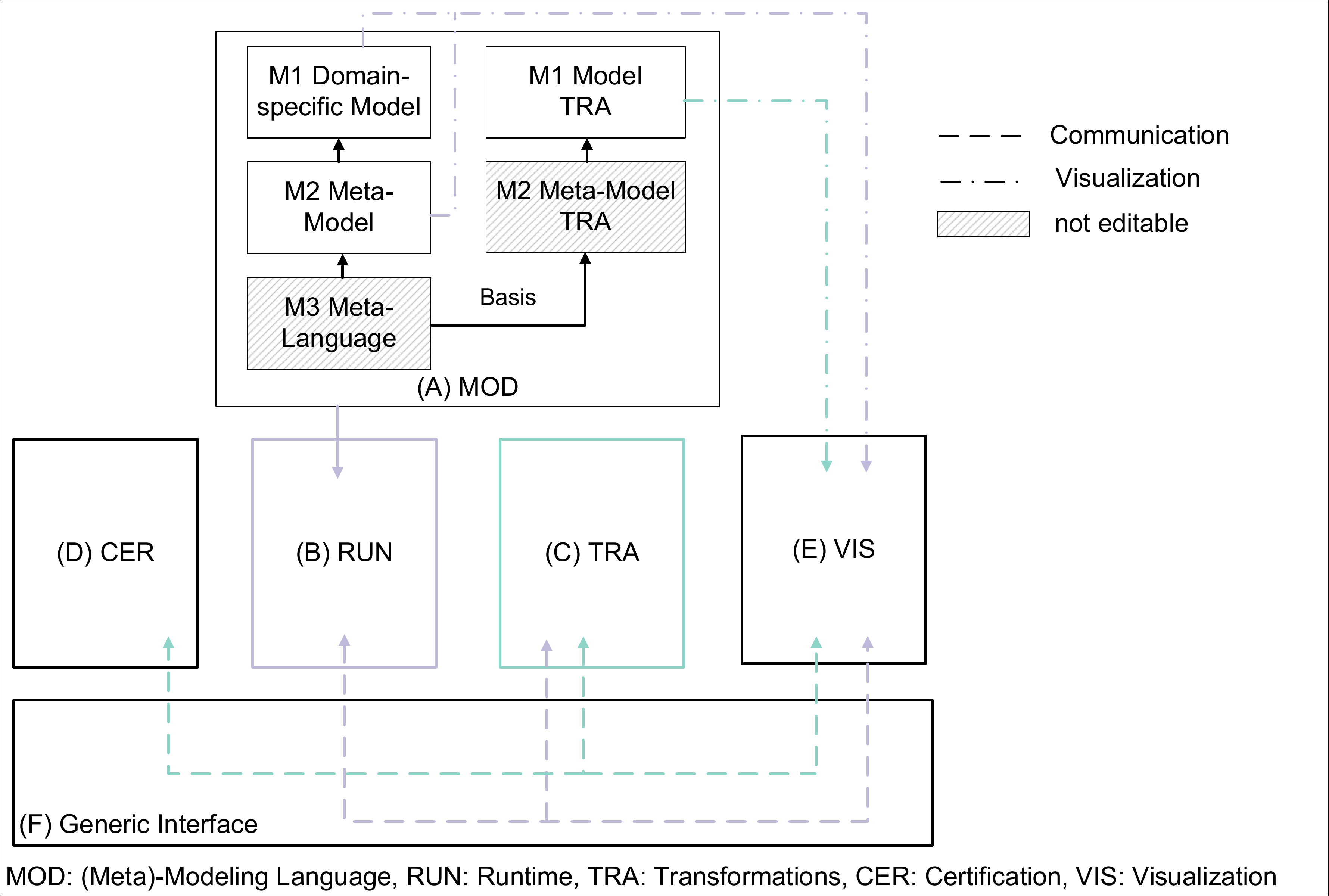}
    \caption{Interaction between Six Cornerstones: A to F}
    \label{fig:interface}
\end{figure}

The realization of the envisioned qualifiable domain-specific modeling framework is currently in progress. In this section, we present first results as well as achievements from prototype implementations of the meta-language and the runtime environment. In Figure \ref{fig:interface}, the software architecture of the six different cornerstones is depicted. The letters (A) to (F) refer to the respective subsection in Section IV. Dashed lines represent the communication over a generic interface (F), and dotted-dashed lines represent the visualization of the different models. The shaded components meta-language and transformation meta-model are not editable. Colors are used to distinguish the (B) RUN part from the (C) TRA part. 
VIS (E) implements a generic graphical visualization. It is a graphical editor for the meta-model and domain-specific model. 
RUN (B) implements MOD (A) and responds to requests from (C) and (E). (C) executes model transformations based on a transformation model, which is based on the M3 meta-language.
The M2 transformation model defines the semantics and enables the creation of various model-to-model transformations. The fact that the transformation meta-model is based on the M3 meta-language optionally allows the utilization of (E) for editing transformations. Via Generic Interface (F) the transformation framework can access meta-level information that is necessary for executing model transformations. For certification purposes, CER (D) automatically generates qualification artifacts based on (C) and (A). The cornerstones (B) to (E) can only communicate over (F).
This simplifies the qualification process due to the possibility of incrementally qualifying the cornerstones. 

\subsection{Development Meta-Language} 
\begin{figure}[!t]
    \centering
    \includevisio[width=0.37\textwidth]{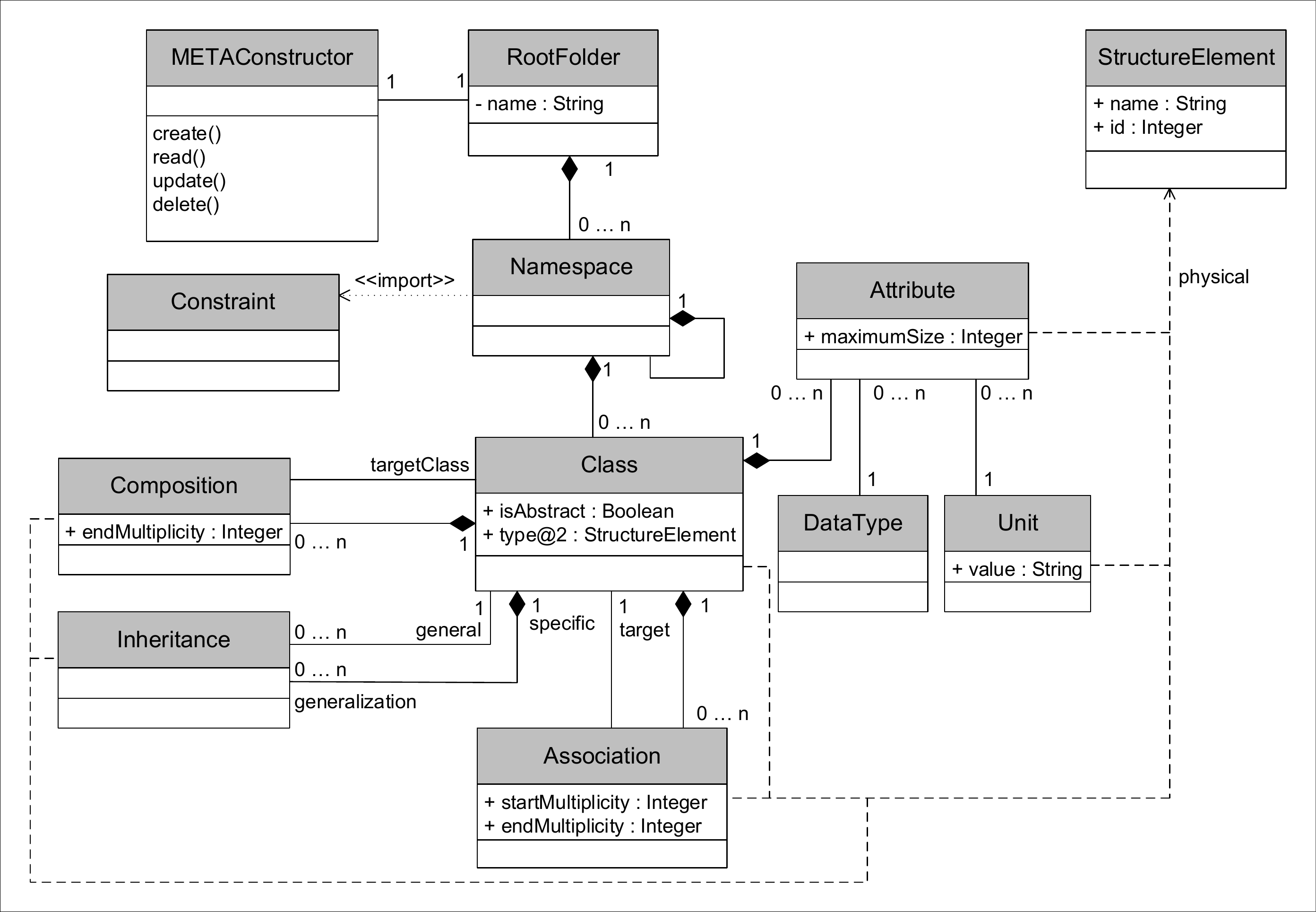}
    \caption{Modeling of Meta-Language (MOD)}
    \label{fig:UML}
\end{figure}

The core of our framework is the meta-language depicted in Figure \ref{fig:UML}. The basis of our meta-language are components from the UML standard like blocks for the representation of classes, composition arrows to indicate  parent-child relations, and solid lines representing associations between classes. Navigability is always possible in both directions. Therefore, the typical representation of associations with arrows is not needed. The UML standard is not sufficient to fit our needs for describing the meta-language and the intended functionality. This includes the requirement to be able to define values at M3 level that must necessarily be set at M1 level. Hence, deep instantiation with potency values and physical classification is utilized from \cite{DeepInstantiation}.\\
The root element of every model is the \textit{RootFolder}. In addition to the \textit{RootFolder}, the \textit{METAConstructor} is available at the beginning of every modeling task. 
The \textit{METAConstructor} is used to create, update, read and delete all available elements of each model on all modeling levels. In order to introduce physical classification, the \textit{StructureElement} is needed. 
It signifies that every element, no matter at what level, is a \textit{StructureElement} with its attributes: \textit{id} and \textit{name}. 
This allows the user to easily change the type of already created objects on M1-level.
In order to structure the meta-model, namespaces are introduced. 
\textit{Namespace} objects are contained by the \textit{RootFolder} as well as the \textit{Namespace} itself.
The \textit{Namespace} is comparable to the package of Ecore \cite{EMF}. The \textit{Constraint} environment is located at the namespace, so constraints can be applied to the whole meta-models or to individual elements within a \textit{Namespace}. A detailed concept of the \textit{Constraint} environment is part of further research activities. The main element of our meta-language is the \textit{Class} which mainly provides the same functionalities as in other modeling languages. The \textit{Class} is extended with the \textit{type} attribute which can be instantiated on M2 level as well as on M1 level, marked with the @ symbol indicating a potency value as described in Section II. A \textit{Class} contains \textit{Attributes}. Every \textit{Attribute} acts as a list of values with at least one entry. As in other modeling-languages, every \textit{Attribute} has a \textit{DataType}. In contrast to other modeling approaches the concept of a \textit{Unit} is introduced. 
With \textit{Units}, we introduce an additional check mechanism when performing transformations and create additional information for the documentation of the meta-model. Every \textit{Class} can be directly connected to at least one of three different types of linkages, (1) an \textit{Association}, (2) a \textit{Composition}, or (3) an \textit{Inheritance}. Three separated linkage types are justified by safer implementations of the linkages in the runtime environment. Implementing universal linkages often requires instance-of operators resulting in performance and qualification issues. The \textit{Composition} describes a parent-child relation. With \textit{Inheritances}, \textit{Classes} can, but do not have to, be defined as abstract. Additionally, multiple \textit{Inheritance} is enabled.
In contrast to other meta-languages the scope of our meta-language is simplified in order to fulfill essential tasks, e.g. the proxy functionality, and annotations are not considered.  

\subsection{M3-model Runtime with Ada}
A first prototype of the meta-language is implemented with Ada, a programming language designed to be used in safety-critical applications \cite{Barnes2015}. The prototype provides the creation of meta-models and domain-specific models via command-line input. Even if it is possible to perform object-oriented programming with Ada, the implementation of the meta-language is performed with an object-relational mapping approach. Every object of the M3 level has its own global list of objects, e.g. a list of all classes, all attributes, all inheritances etc. The interaction of the object-relational approach and the global availability of the list facilitates avoiding the usage of pointers. In each of these lists, the mandatory values are stored as well as references, with global unique identifiers, to other objects.
One advantage of the object-relational mapping over the object-oriented approach is that  e.g. it is possible to easily change the type of a classical object during runtime.

Every input command is based on a CRUD (create, read, update, delete) request provided with an identifier which can globally identify every object. The read and update request additionally needs information about the processed attribute of an object. For the update request, additionally, the new value of the attribute is required.

With the exception of the \textit{Constraint} environment and the \textit{RootFolder}, the current demonstrator covers the entire meta language. To further reduce the qualification effort, it is currently investigated whether meta-languages can be implemented with the strongly restrictive programming language SPARK \cite{SPARK}, a subset of Ada. 

\section{Conclusion and Outlook}
Within this paper, we introduced a concept and requirements of a qualifiable (meta)-modeling framework suitable for the development of safety-critical systems. The framework consists of six independent cornerstones, (1) the essential meta-language, (2) a minimalistic, qualifiable runtime, (3) deterministic, verifiable, (graphical) transformations, (4) the qualification artifact generation, (5) a decoupled visualization, (6) and the decoupled interaction of framework components. 
All mentioned aspects shall support the user in the tool and system qualification process, while providing an up-to-date user experience. This is achieved by e.g. the possibility of an independent qualification of the different aspects due to the generic interface and by implementing the runtime with the programming language Ada (SPARK).
Within our framework some research tasks were already conducted like the development of the meta-language and the implementation of a first runtime prototype. The next research tasks include the implementation of: (i) a qualifiable, deterministic transformation language which shall be based on MOD; (ii) a concept for qualification artifact generation (traceability features of model changes, documentation, etc.); and (iii) a browser-based generic visualization.

\section*{Acknowledgment}

The German Federal Ministry for Economic Affairs and
Energy (BMWi) has funded this research within the LUFO-VI program and the TALIA project.

\newpage
\bibliographystyle{IEEEtran}
\bibliography{IEEEabrv,vision_paper}

\end{document}